\documentclass[final, 5p, number,sort&compress, twocolumn]{elsarticle}
\usepackage{lineno}
\journal{NIM-A}

\usepackage{amsmath}
\usepackage{amssymb}
\usepackage{subcaption}
\usepackage[utf8]{inputenc}
\usepackage{color}
\usepackage{url}
\usepackage{graphicx} 
\usepackage{textcomp}
\usepackage{siunitx}
\usepackage{gensymb}
\usepackage{csquotes}
\usepackage{xcolor}
\renewcommand{\mkbegdispquote}[2]{\itshape}
\usepackage{clipboard}
\usepackage[export]{adjustbox}[2011/08/13]
\usepackage[framemethod=tikz]{mdframed}
\usepackage{listings}
\usepackage{comment}
\usepackage{etoolbox}
\newbool{hasReply}
\setbool{hasReply}{false}

\begin{document}
\begin{frontmatter}
\title{Space Charge Effects in the S$\pi$RIT Time Projection Chamber}
\author[1,2]{C.Y. Tsang\corref{cor1}}
\ead{tsangc@nscl.msu.edu}
\author[1,2]{J. Estee}
\author[1]{R. Wang}
\author[1,2]{J. Barney}
\author[1]{G. Jhang}
\author[1,2]{W.G. Lynch\corref{cor1}}
\ead{lynch@nscl.msu.edu}
\author[1]{Z.Q. Zhang}
\author[1]{G. Cerizza}
\author[3]{T. Isobe}
\author[4]{M. Kaneko}
\author[3]{M. Kurata-Nishimura}
\author[5]{J.W. Lee}
\author[4]{T. Murakami}
\author[1,2]{M.B. Tsang}
\author{and the S$\pi$RIT collaboration}


\address[1]{National Superconducting Cyclotron Laboratory, East Lansing, MI 48824 USA}
\address[2]{Department of Physics and Astronomy, Michigan State University, East Lansing, MI 48824 USA}
\address[3]{RIKEN Nishina Center, Hirosawa 2-1, Wako, Saitama 351-0198, Japan}
\address[4]{Department of Physics, Kyoto University, Kita-shirakawa, Kyoto 606-8502, Japan}
\address[5]{Department of Physics, Korea University, Seoul 02841, Republic of Korea}

\cortext[cor1]{Corresponding author}

\begin{abstract}
Time projection chambers (TPCs) are widely used in nuclear and particle physics. They are particularly useful when measuring reaction products from heavy ion collisions. Most nuclear experiments at low energy are performed in a fixed target configuration, in which the unreacted beam will pass through the detection volume. As the beam intensity increases, the buildup of positive ions created from the ionization of the detector gas by the beam creates the main source of space charge, distorting the nominal electric field of the TPC. This has a profound effect on the accuracy of the measured momenta of the emitted particles. In this paper we will discuss the magnitude of the effects and construct an observable appropriate for fixed target experiments to study the effects. We will present an algorithm for correcting the space charge and discuss some of the implications it has on the momentum determination. 
\end{abstract}


\end{frontmatter}

\section{Introduction}
\label{Intro}

When placed inside of a magnetic field, time projection chambers (TPCs) provide an accurate means to determine the momenta of charged particles resulting from particle collisions. TPCs in nuclear physics experiments are typically used in either fixed-target or active-target mode~\cite{Koshchiy2019, AYYAD2018, EOSTPC1990, Shane2015}. Charged particles traversing the gas-filled detection volume of a TPC ionize the gas, producing electron-ion pairs. The ions are much heavier than the electrons. As a result, the ions drift much slower in the TPC and a build-up of ions can occur within the TPC. One must take into consideration the effects that the build up of positive free ion, known as space charge, have on the reconstructed tracks of the charged particles. 

The gas ionization (and resulting space charge) occurs due to several processes: from charge amplification within the avalanche region, from the reaction products of nuclear reactions, and from unreacted beam particles traversing the detection volume. The space charge originating from the back flow of ions from the avalanche region is mitigated by the use of a gating grid~\cite{Tangwancharoen2016}, which, when operated with the proper settings, blocks at least 99.9\% of back flow. Typically, the frequency of nuclear reactions is small compared to the rate of incoming beam particles; for the data examined in this work, nuclear reactions only occur from about 2\% of the incident beam particles. Further, the products from a nuclear reaction are mainly composed of protons and other light particles that have a low stopping power as compared to the heavy beam. For example, a LISE++~\cite{Tarasov2008} calculation shows that the ratio of energy deposition per unit length in P10 gas (a typical gas used in TPCs, consisting of 90\% Ar and 10\% CH$_4$) of a \SI{500}{MeV/c} proton to that from a single Sn beam at \SI{270}{MeV/u} is only 0.03\%. Due to the relatively low rate of occurrence and low ionization, the contribution to space charge from nuclear reactions is diminutive. Therefore, in this paper we consider only the space charge accumulation due to unreacted beam particles traversing the TPC detection volume.

Our studies specifically analyze the space charge effects observed within data taken by the SAMURAI Pion-Reconstruction and Ion-Tracker (S$\pi$RIT) TPC during experiments performed at the Radioactive Isotope Beam Factory (RIBF) facility at RIKEN in Japan~\cite{Shane2015}. The S$\pi$RIT TPC is a rectangular TPC with an active gas volume of dimensions \SI{86.2}{cm} $\times$ \SI{51.3}{cm} $\times$ \SI{134.4}{cm}, constructed in a joint US-Japan collaboration~\cite{Shane2015,Barney2019}. The pad-plane, which detects the position of the drift electrons, is located on top of the active gas volume while the cathode, which removes the associated positive ions, is located below the gas volume. A set of experiments were performed in 2016 with four secondary Sn isotope beams, $^{108}$Sn, $^{124}$Sn, $^{112}$Sn and $^{132}$Sn at an energy of 270 MeV/u. The beam intensity varied but was roughly on the order of \SI{10}{kHz}. The beams impinged on a target located \SI{13.2}{mm} upstream of the TPC's detection region, with the $^{112}$Sn target being used for the first two beams and the $^{124}$Sn target for the last two.

An ideal way to directly measure the space charge effects is to use a laser system. Highly energetic photons from the laser ionize gaseous molecules, which are detected by the TPC. The deviation of the detected trajectory from a straight line provides a measurement of the distortion caused by space charge and B-field non-uniformity~\cite{Mooney2015}. The S$\pi$RIT TPC did not have a functioning laser calibration system during this experimental campaign. 

The STAR TPC collaboration showed that the amount of space charge is sensitive to observables such as the signed Distance of Closest Approach (sDCA)~\cite{Star2002}. This is suitable for collider experiments where reactions occur everywhere along the beam path. As the desired nuclear reactions occur only on the target plane in the S$\pi$RIT experiment, a new observable, $\Delta V_{LR}$, is constructed to correctly reflect this geometric configuration. It will be used in the development of the correction algorithm for the space charge distortion in this study.

\section{Experimental setup}
\label{exp_set_up}
The S$\pi$RIT TPC was designed to measure the yields of charged pions and light charged particles in heavy ion collisions, an observable which is sensitive to the symmetry energy term of the nuclear equation of state (EoS)~\cite{Shane2015}. In this paper, we choose to use data from the reaction \mbox{$^{108}$Sn + $^{112}$Sn} to illustrate the effects of the space charge. This set of data is chosen because the beam intensity varies more than the other beams used in the experiment, providing a better demonstration of the correlation between space charge effects and beam intensity.  The same corrections and conclusions should be applicable to the other reactions as well. 

\begin{figure}[!h]
\includegraphics[width=1\linewidth, trim=0cm 0cm 0cm 0cm, clip=true]{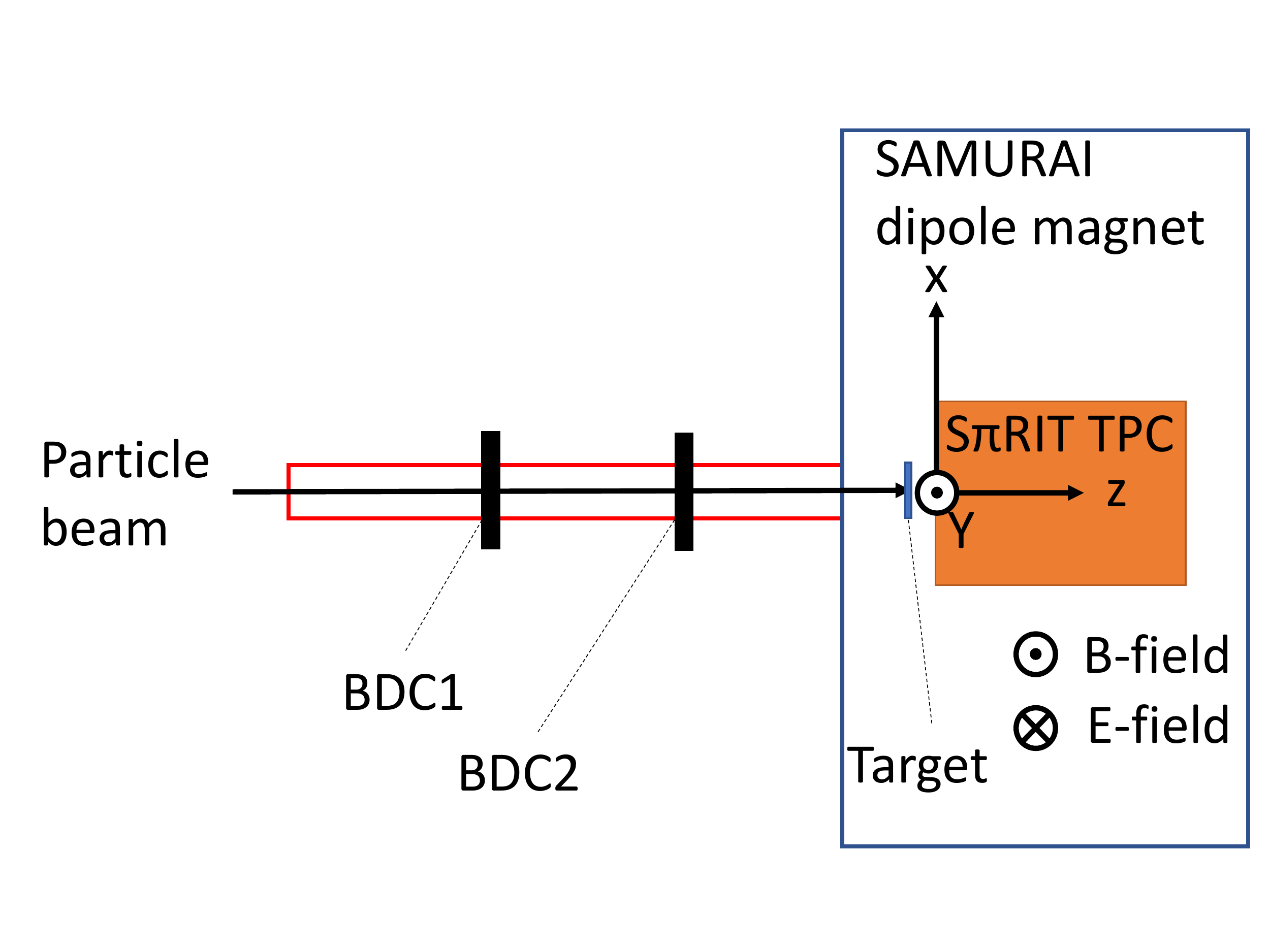}
\caption{Schematic drawing of the experimental setup when viewed top-down. The beam inside the beam pipes enters the TPC from the left. The beam-tracking detectors, BDC1 and BDC2, are labeled. The origin of the coordinate system and the direction of the fields are illustrated, with y-direction pointing out of the page}
\label{setup}
\end{figure}

A top-down view of the experimental setup is shown in Fig.~\ref{setup}. The S$\pi$RIT TPC is installed inside the SAMURAI dipole magnet~\cite{SAMURAI}, which provides a nearly uniform \SI{0.5}{T} magnetic field. The detection volume of the TPC is enclosed by the field cage, which is made of printed circuit boards with precisely spaced horizontal copper strips. The voltage of each strip decreases sequentially from top to bottom to create a uniform electric field of \SI{124.7}{V/cm}~\cite{Changj2016}. Figure~\ref{setup} also defines the coordinate system for this study, with the origin $(0,0,0)$ located at middle of the front edge of the active pads and at the same height as the pad-plane. The magnetic field points in the $+y$ direction while electric field points in the $-y$ direction; the $+z$ direction points along the beam defining a right handed coordinate system.

The TPC detection volume is filled with P10 gas at atmospheric pressure which is continuously monitored. There are several detectors upstream of the target, including two Beam Drift Chambers (BDC1 and BDC2) that track the beam as it passes through the beam tube~\cite{Kobayashi2013}. By extrapolating the beam trajectory to the target plane at ${z=\SI{-13.2}{mm}}$,  the reaction vertex can be determined with position resolution of ${\delta x=\SI{0.7}{mm}}$ and ${\delta y=\SI{0.3}{mm}}$~\cite{Barney2019}. Although the reaction vertex can be and is usually reconstructed from the tracks in the TPC~\cite{RAVE}, it is susceptible to space charge effects which renders it unusable in this study. Throughout this paper, ``vertex'' refers to the vertex extrapolated to the target plane by the two BDC measurements. The incident beam intensity is calculated through the data analysis of other upstream detectors (not shown in Fig.~\ref{setup}). Details of this analysis can be found in Ref.~\cite{Barney2019}. 

\section{\label{SCD}Space charge distortion}

In the case of anti-parallel E- and B-fields, electrons created by the traversing charged particles will drift opposite to the E-field, terminating on the pad-plane. If the E- or B-fields exhibits any non-uniformities, the electron drift path will not be completely vertical. The reconstructed positions of electron clusters along the trajectory of the primary ionizing particle will be distorted if this lateral movement of electrons is not corrected for. In the presence of E- and B-fields, the motion of electrons is described by the Langevin equation~\cite{Drift1993}, 

\begin{equation}
\frac{d\vec{x}}{dt} = \frac{\mu}{1+(\omega\tau)^2}\Big(\vec{E} + \omega\tau\frac{\vec{E}\times\vec{B}}{\lVert\vec{B}\rVert}+\omega^2\tau^2\frac{\vec{E}\cdot\vec{B}}{\lVert\vec{B}\rVert^2}\vec{B}\Big).
\label{drift}
\end{equation}
In this equation, ${\omega=e\lVert\vec{B}\rVert/m}$ is the cyclotron frequency, $\tau$ is the mean time between collisions, $m$ is the mass and $e$ is the signed charge of the drifting electron or ion and ${\mu=e\tau/m}$ is the ion-mobility. The value of $\tau$ is fixed once the drift velocity is measured. Since the E- and B-fields are nearly uniform, the drift velocity of electrons can be approximated as ${\vec{v} = \mu\vec{E} = e\tau\vec{E}/m}$. By comparing hit points with the longest drift time to the height of the device, the electron drift velocity for S$\pi$RIT with data from $^{108}$Sn beam is determined to be ${\SI{5.42(6)}{cm/\micro\second}}$, with $\omega\tau = 2.18(2)$.

\begin{figure}[!h]
\includegraphics[width=1\linewidth]{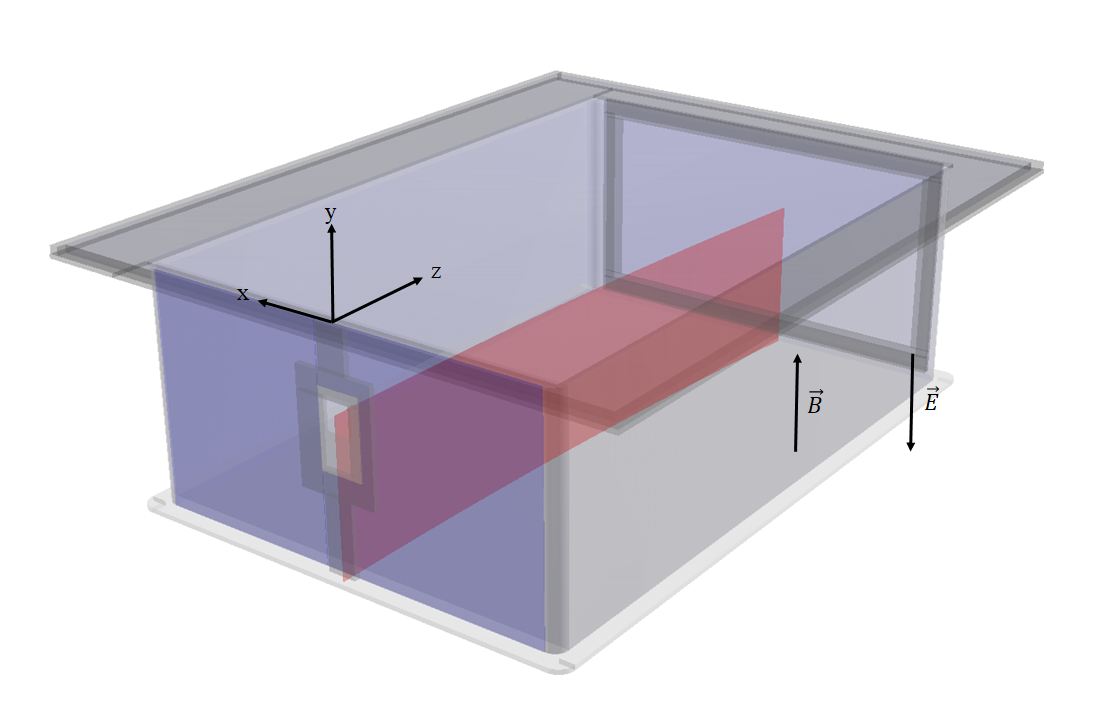}
\caption{Illustration of the approximated space charge density in S$\pi$RIT TPC. The lightly shaded rectangle on the front face is the entrance windows and the highlighted region inside the TPC assumes a uniform space charge distribution created by the incoming beam.}
\label{SCGraph}
\end{figure}

The beam-induced ions drift primarily downward towards the cathode with uniform velocity. While the repulsion between ions causes this sheet to diffuse slightly as they drift downward, it has a small effect on the electron trajectories. Therefore, they are approximated as a positively-charged sheet as indicated by the shaded region inside the TPC in Fig.~\ref{SCGraph}. 

The energy loss per unit length of $^{108}$Sn at \SI{270}{MeV/u} inside P10 gas at atmospheric pressure is around ${\sim\SI{10}{MeV/cm}}$ and the ion drift velocity is ${\sim\SI{2e-4}{cm/\micro\second}}$.  Given that $\sim30$ electrons are ionized from the gas per eV, a beam intensity of \SI{10}{kHz} would give rise to a sheet charge density on the order of ${\sim\SI{e-8}{C/m\squared}}$. Using the empirical charge-finding method, which will be discussed in Section~\ref{SpaceChargeCorrection}, the accurate value is found to be \SI{2.1e-8}{C/m\square}. Furthermore, the ionization as measured by the energy loss per unit length of the beam particle at the beginning ($z=0$) and end (${z=\SI{134.4}{cm}}$) of the TPC only differs by $2.2\%$ according to LISE++~\cite{Tarasov2008}. Therefore, we assume the sheet charge density along the beam path to be uniform.

\begin{figure}[!h]
\includegraphics[width=0.9\linewidth, clip=true, trim=0.5cm 0.5cm 1.5cm 2cm, clip]{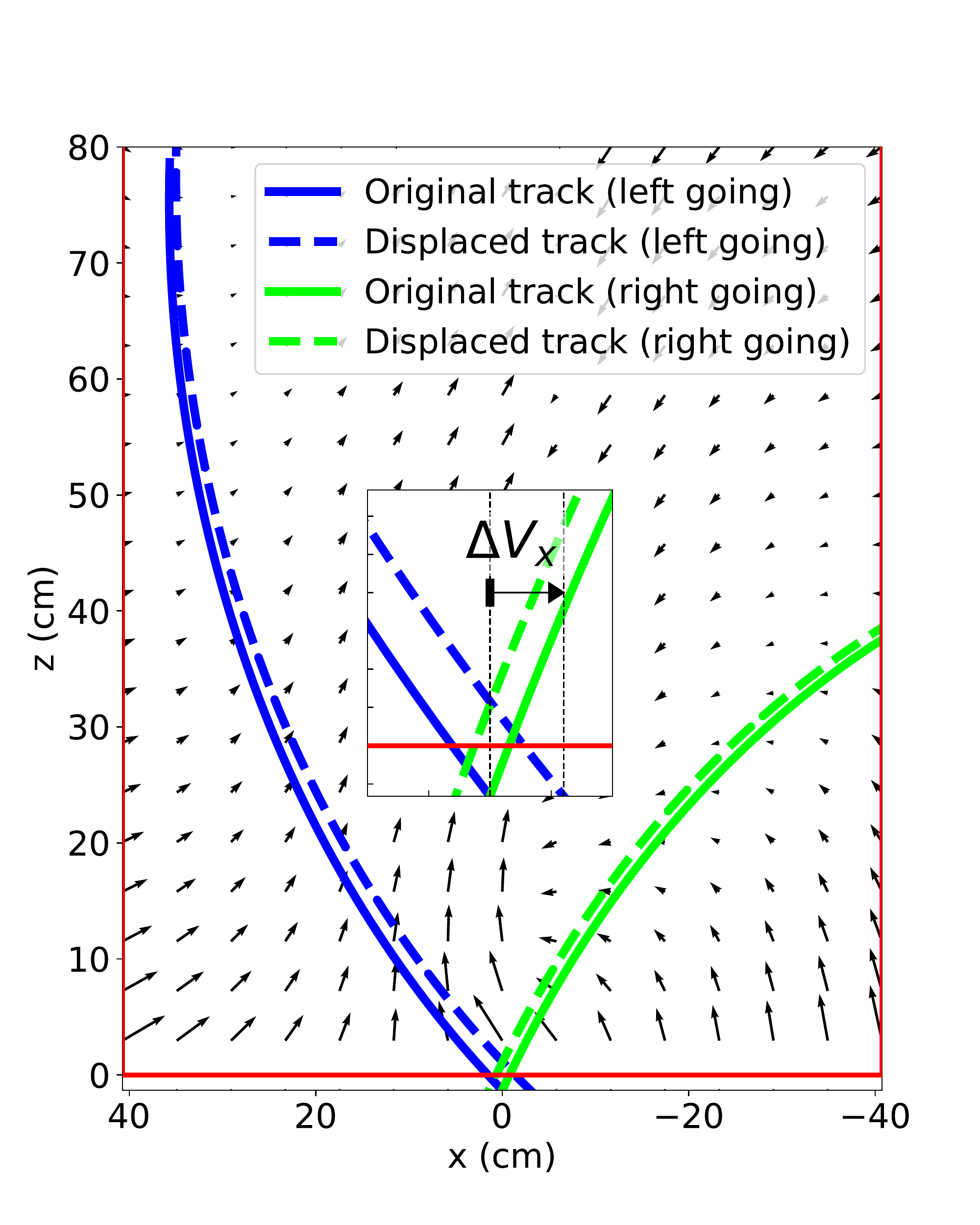}
\caption{Top down view of the TPC tracks as a demonstration of space charge effects, with the vector field representing the space charge displacement of drift electrons. Note that the x-coordinate increases from right to left; this is chosen to reflect the physical orientation of the TPC when viewed from top to bottom. The solid lines represent the original locations of the drift electrons while the dashed lines represent the shifted locations when electrons reach the pad-plane. The inset in the center illustrates the geometric meaning of $\Delta V_{x}$. The large rectangle starting at $z=0$ indicates the active detection volume of the TPC and the reaction vertex is located on the target plane at $z= -13.2\ \text{mm}$.}
\label{SC_Effect}
\end{figure}

To understand the movement of electrons, let us consider only the electric field contribution from the sheet charge; the first term ($\vec{E}$ term) inside the parenthesis of Eq.~\eqref{drift} pushes electrons
in the direction normal to the sheet charge. The second term (the $\vec{E}\times\vec{B}$ term) pushes electrons in the direction parallel to the beam path. The third term (the $\vec{E}\cdot\vec{B}$ term) pushes electrons in the direction of B-field. The influence the space charge has on the trajectories is a complicated three-dimensional effect due to the non-uniformity of SAMURAI 
magnetic field and the space charge configuration. The details on obtaining a solution from Eq.~\eqref{drift} and measuring the sheet charge density will be discussed in later sections. However, general tendencies can be illustrated with an example of the displacement vector field calculated from Eq.~\eqref{drift} with a typical sheet charge density of \SI{2.1e-8}{C/m\squared} (corresponds to beam intensity of \SI{10}{kHz}), as shown in Fig.~\ref{SC_Effect} for drift electrons originating from ${y=\SI{-25}{cm}}$. The magnitude of the vectors are scaled up for illustration purposes. To better demonstrate its effect, two sets of example tracks for positively-charged particles are drawn, with the two tracks on the right of the vertex traveling in $-x$ direction (right-going), and the other two traveling in $+x$ direction (left-going). The original undistorted tracks are labeled by the solid lines and the displaced tracks are given by the dashed lines. Negatively charged particles, consisting of $\pi^-$ and $e^-$, are not included in the illustration for brevity, but are affected in a similar way by space charge effects.

The distortion causes the intersection point between the reconstructed tracks and the target plane at ${z=\SI{-13.2}{mm}}$ to diverge from the actual vertex. The difference in the $x$-coordinate between the vertex and the projected position on the target plane is denoted as $\Delta V_{x}$. This is highlighted by the inset in Fig.~\ref{SC_Effect}, where the $\Delta V_{x}$ of the displaced left-going track is shown. Generally speaking, the space charge effects cause the projected positions of left-going tracks to be shifted to the right (negative $\Delta V_{x}$), and for right-going tracks to be shifted to the left (positive $\Delta V_{x}$). It should be noted that although Fig.~\ref{SC_Effect} shows a vertex at $x=0$, the measured coordinates of the vertex changes event-by-event.

\begin{figure}[!h]
 \includegraphics[width=1\linewidth]{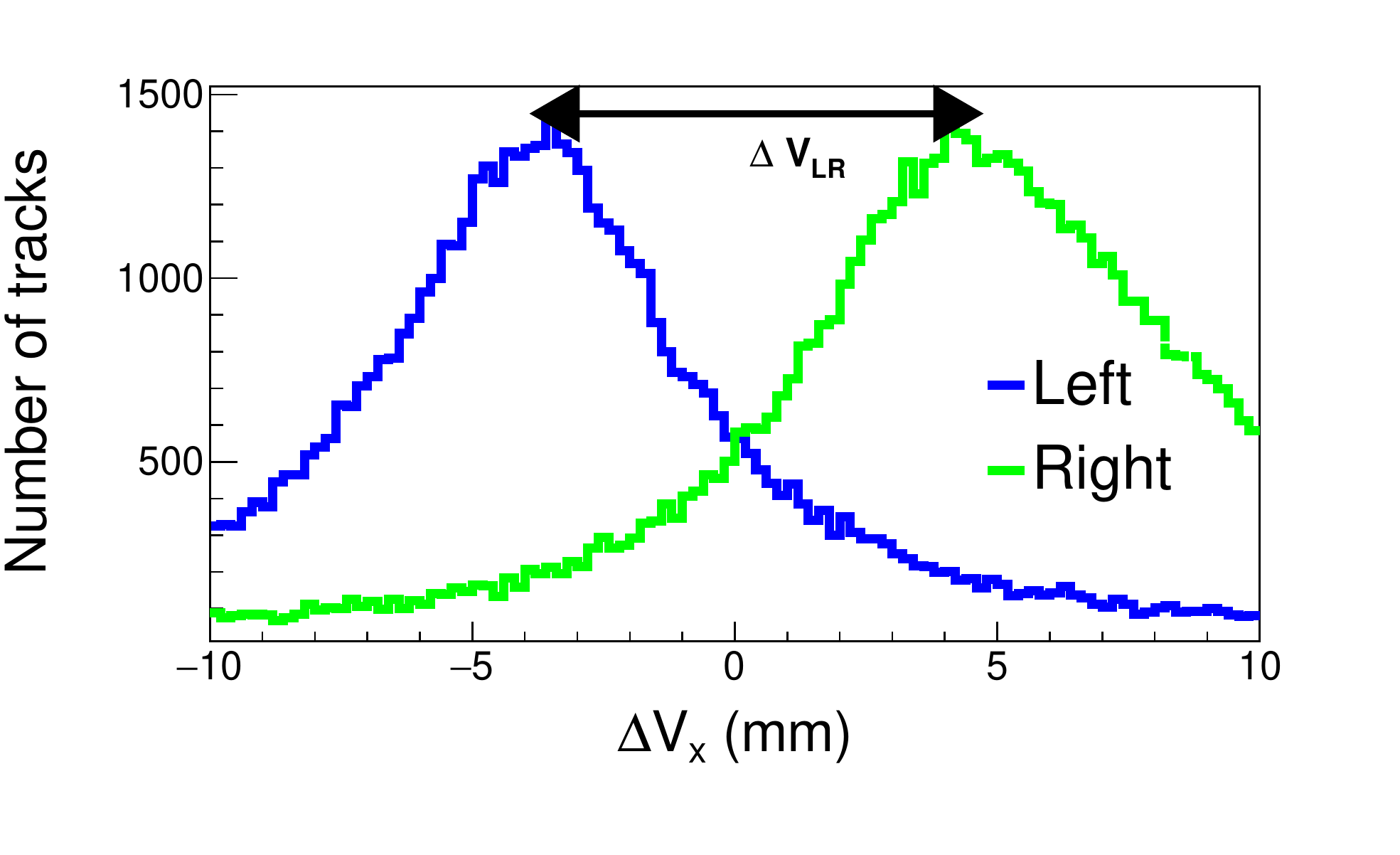}
\caption{Distribution of $\Delta V_{x}$ for particles with polar angle larger than 40\degree, separating particles heading in the $+x$ (left) and $-x$ (right) directions. The separation between the two peaks is defines as $\Delta V_\text{LR}$.}
 \label{DVTP_raw}
\end{figure}

We observe a similar trend in the data; Fig.~\ref{DVTP_raw} shows two $\Delta V_{x}$ distributions for left- and right-going tracks for particles emitted with polar angle $\theta$, defined with respect to the $z$-axis, larger than 40\degree. The trend is the same for tracks emitted with $\theta < 40\degree$, but to a lesser extent because the electrons are mostly displaced in the $z$-direction, as evident in Fig.~\ref{SC_Effect}. When the tracks are projected onto the target plane, the shift in $z$-coordinate is difficult to observe for tracks that are almost perpendicular to the target plane. The separation between the peak locations of the left- and right-going distributions is denoted as $\Delta V_\text{LR}$. Imperfections in track reconstruction, such as the appearance of broken tracks, cause the long tails in the distributions.

\begin{figure}[!h]
 \includegraphics[width=1\linewidth]{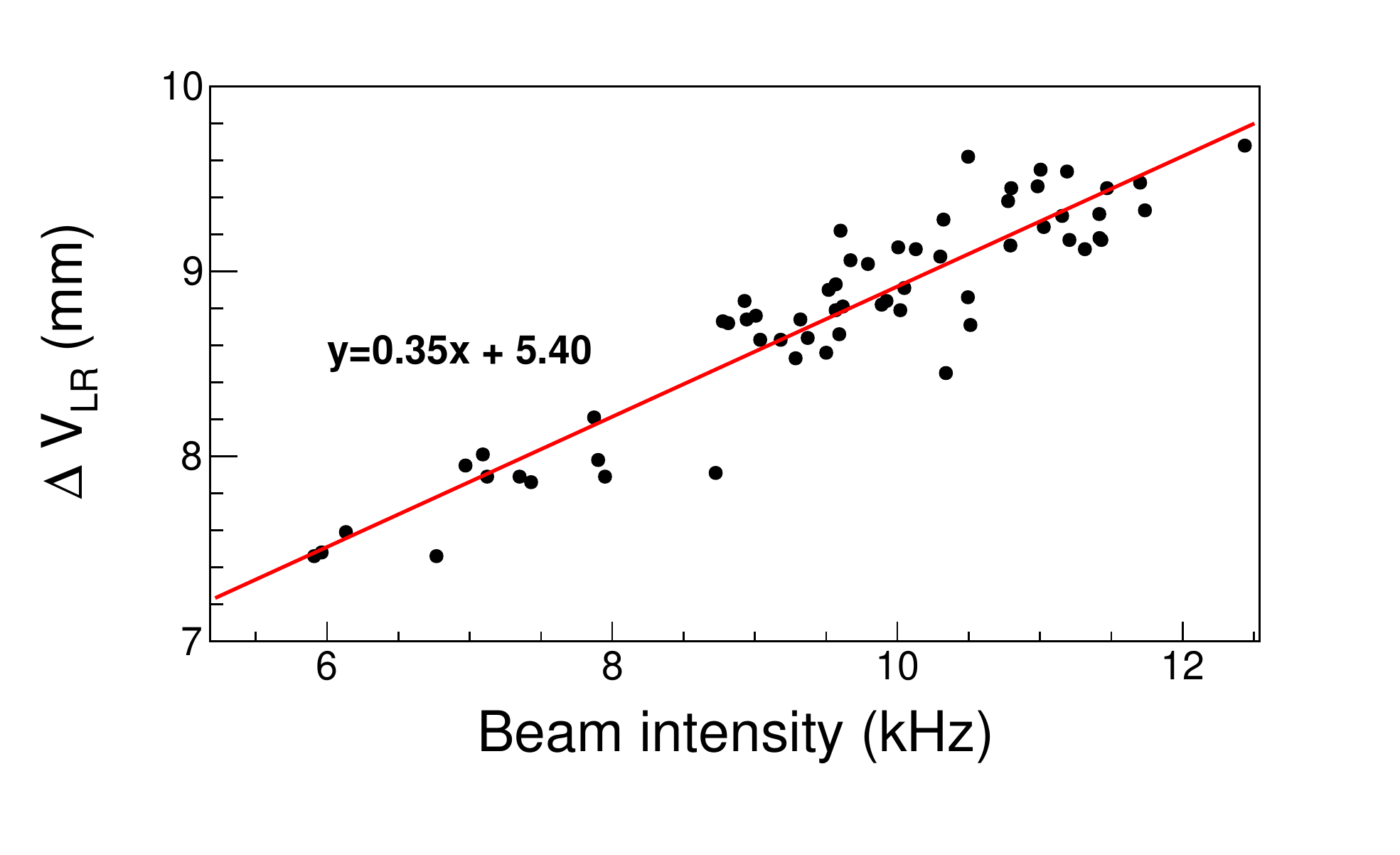}
 \caption{$\Delta V_\text{LR}$ plotted against beam intensity.}
 \label{Sep_V_BR}
\end{figure}

Since the amount of space charge is proportional to the beam intensity, we should 
observe a positive correlation between $\Delta V_\text{LR}$ and the beam intensity if the displacement from the vertex is indeed caused by space charge distortion. During the experiment, data was taken every 30 minutes and constitutes a ``run". The run-averaged value of $\Delta V_\text{LR}$ is plotted against the run-averaged beam intensity in Fig.~\ref{Sep_V_BR} using data from 61 consecutive runs with the $^{108}$Sn beam; a strong linear positive correlation is observed. In the absence of space charge effects, the non-uniformities of the magnetic field alone can cause distortions explaining the non-zero $y$-intercept value of \SI{5.4}{mm}.

The distortions in the measured hit points impact the accuracy of the momentum determination of the emitted particles, where most information relevant to
nuclear physics is encapsulated~\cite{Li2002, Hong2013}. In theory, fitting the track curvatures while including the precise vertex point determined by the BDCs should improve the momentum resolution~\cite{blum2008particle}. In 
practice, the TPC hit points are affected by the space charge while the vertex point determined by the BDCs is not. The combination of 
distorted and undistorted information adversely affects the track fitting algorithm.

\begin{figure*}[!h]
\begin{subfigure}[b]{0.5\linewidth}
 \includegraphics[width=1\linewidth, trim=1.5cm 1cm 0.9cm 1cm, clip=true]{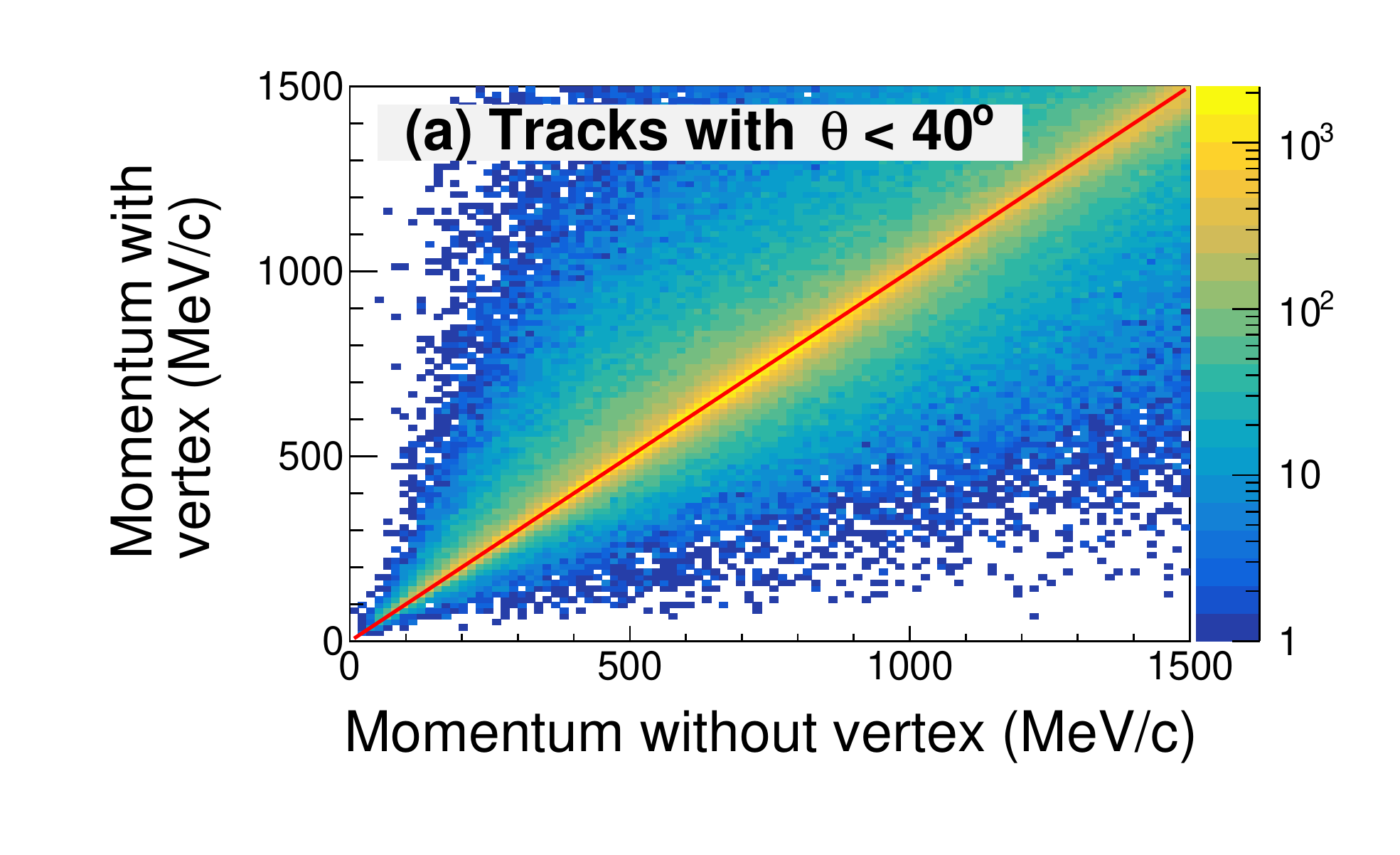}
\end{subfigure}
\begin{subfigure}[b]{0.5\linewidth}
 \includegraphics[width=1\linewidth, trim=1.5cm 1cm 0.9cm 1cm, clip=true]{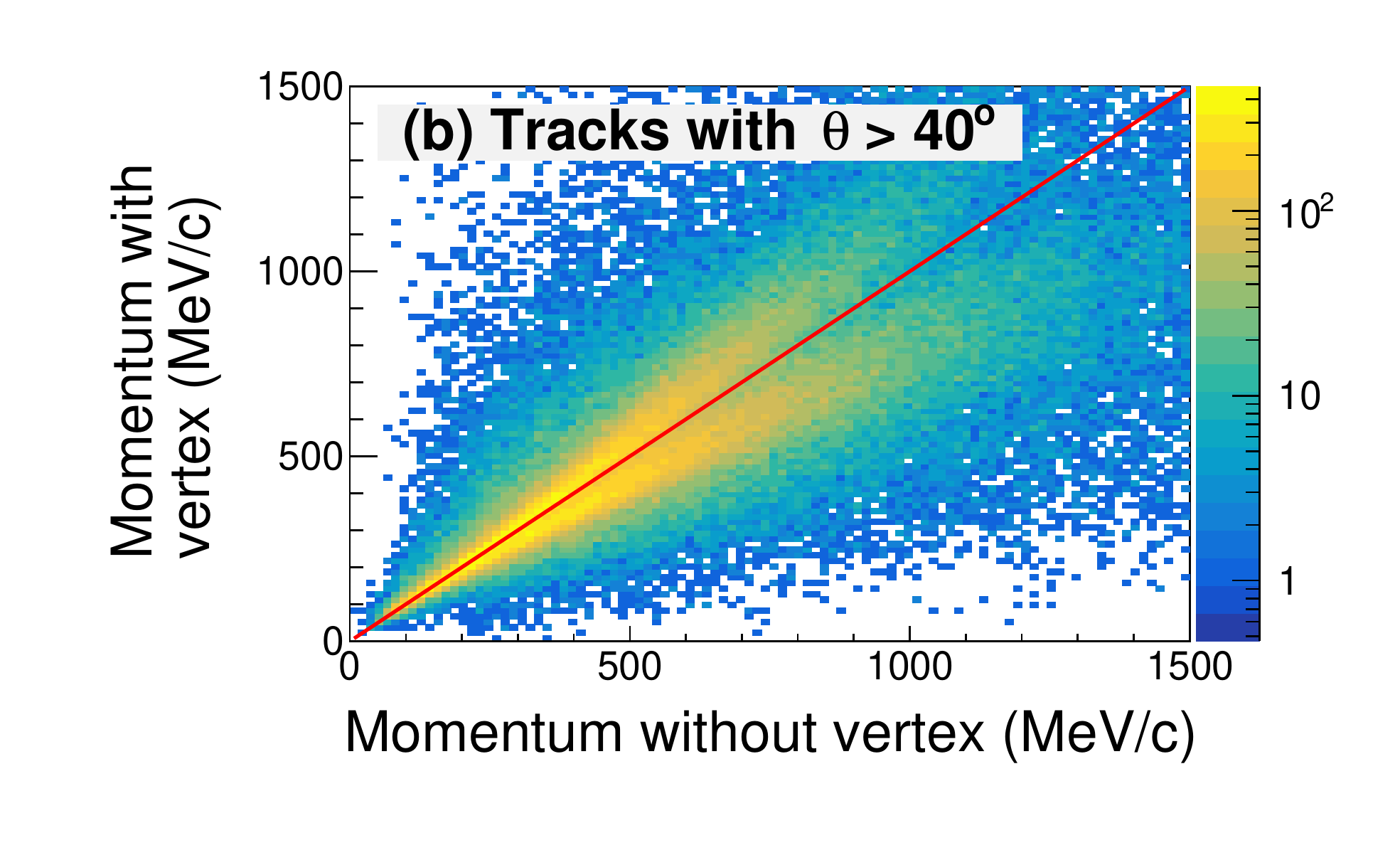}
\end{subfigure}
\begin{subfigure}[b]{0.5\linewidth}
 \includegraphics[width=1\linewidth, trim=1.5cm 1cm 0.9cm 0.5cm, clip=true]{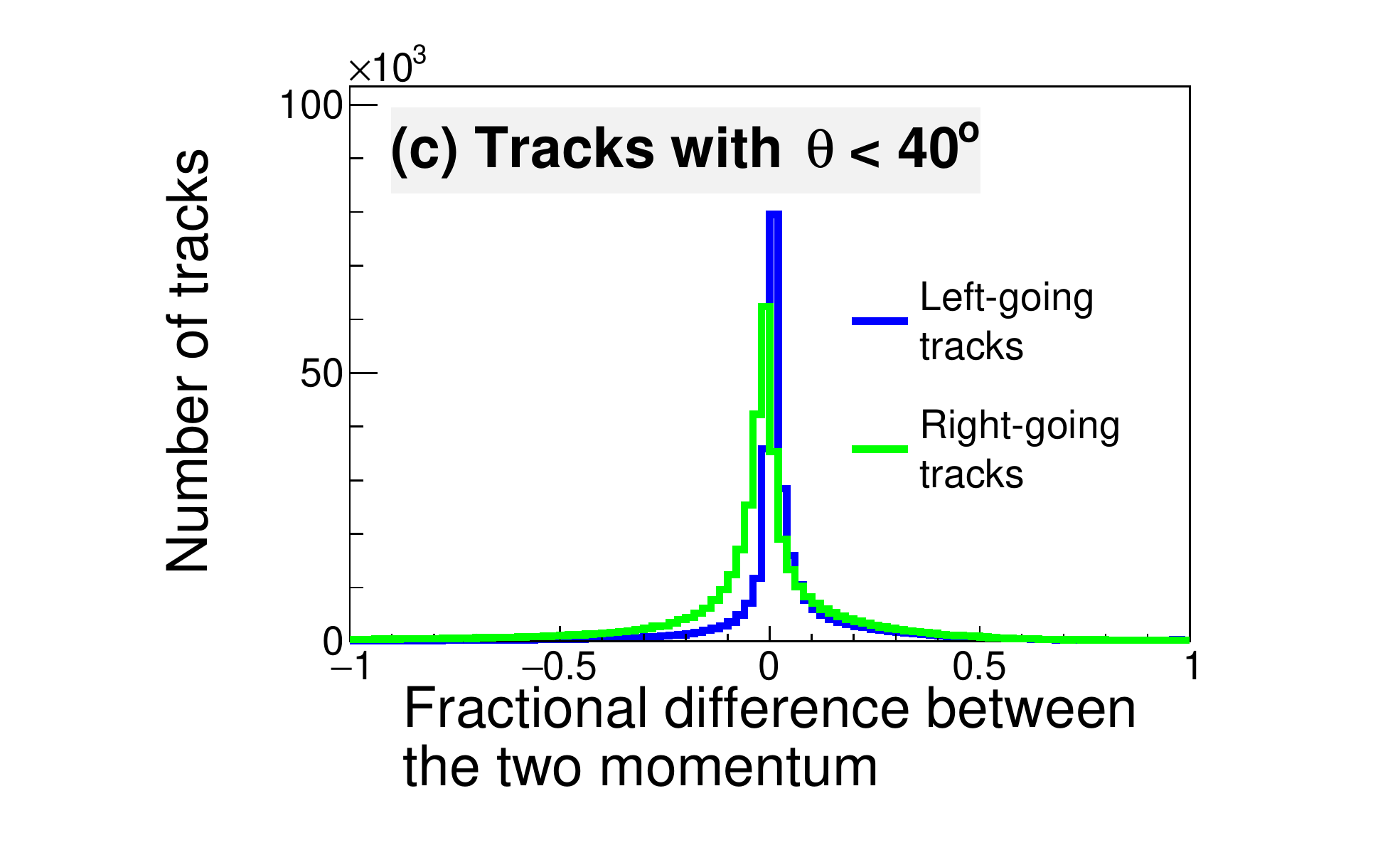}
\end{subfigure}
\begin{subfigure}[b]{0.5\linewidth}
 \includegraphics[width=1\linewidth, trim=1.5cm 1cm 0.9cm 0.5cm, clip=true]{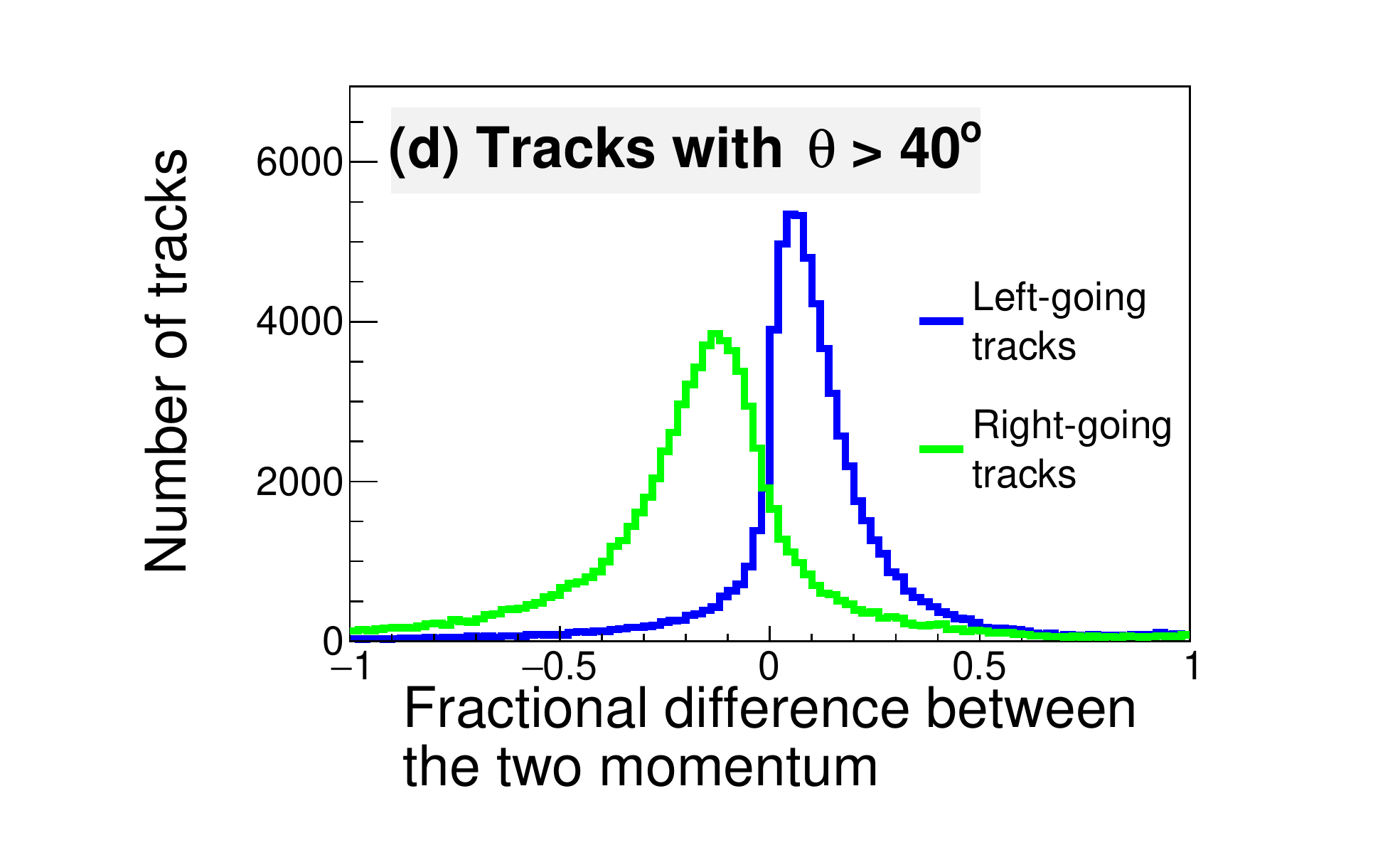}
\end{subfigure}
\caption{The reconstructed momentum including the vertex point is plotted against the reconstructed momentum without the vertex point, for (a) tracks with $\theta$ $<$ 40\degree and (b) tracks with $\theta$ $>$ 40\degree. The color scale represents the number of tracks in each bin. Fractional difference between momentum with and without vertex is also plotted, for (c) tracks with $\theta$ $<$ 40\degree and (d) tracks with $\theta$ $>$ 40\degree. Distribution from left-going tracks and right-going tracks are overlaid in (c) and (d) to demonstration the origin of the bimodal distribution.}
\label{BDCP}
\end{figure*}

The two-dimensional plots in Fig.~\ref{BDCP}a and Fig.~\ref{BDCP}b show the reconstructed momentum with and without the vertex point for tracks with ${\theta < 40\degree}$ and ${\theta > 40\degree}$ respectively. The histograms in Fig.~\ref{BDCP}c and Fig.~\ref{BDCP}d show the fractional difference between the two reconstructed momenta, again for tracks with ${\theta< 40\degree}$ and ${\theta > 40\degree}$ respectively; with the left-going tracks and right-going tracks being plotted separately.
The disagreement before and after refitting with the vertex becomes larger as the polar angle of the track increases since the track's projection onto the target is sensitive to even small distortions. The bifurcation seen in Fig.~\ref{BDCP}b originates from the asymmetry of left- and right-going tracks as discussed earlier and shown in Fig.~\ref{DVTP_raw} 
This is corroborated in Fig.~\ref{BDCP}d where the distributions from tracks heading in opposite $x$-directions peak at different locations, with left-going tracks peaking at 0.054 and right-going tracks peaking at -0.14.

\section{\label{SpaceChargeCorrection}Space charge correction}
In this section, we will discuss in detail the algorithm for correcting the space charge distortions. We define $x_\text{det}$ and $z_\text{det}$ as the $x$- and $z$-coordinates of the measured cluster positions on the pad-plane, and $x_\text{real}$ and $z_\text{real}$ as the $x$- and $z$-coordinates of the original position of the ionization. An inverse map is defined as the difference in coordinates between the detected and original position,

\begin{equation}
\begin{split}
\Delta x(x_\text{det}, z_\text{det}, t_\text{drift}) = x_\text{real}(x_\text{det}, z_\text{det}, t_\text{drift}) - x_\text{det},\\
\Delta z(x_\text{det}, z_\text{det}, t_\text{drift}) = z_\text{real}(x_\text{det}, z_\text{det}, t_\text{drift}) - z_\text{det}. \\
\end{split}
\label{invmap}
\end{equation}
The variable $t_\text{drift}$ is the drift time of the electrons, which is measured as the time difference between the associated pad signal and the trigger. With accurate maps of the E- and B-fields, the inverse map can be calculated by solving Eq. \eqref{drift} with Runge-Kutta methods; by setting $x_\text{det}$, $z_\text{det}$ and $t = t_\text{drift}$ as the initial position, with negative time steps, we solve and trace the electrons back to where they originated at $t = 0$, corresponding to $x_\text{real}$ and $z_\text{real}$ in Eq.~\eqref{invmap}. Tables of $\Delta x$ and $\Delta z$ are generated in this manner for a range of values in $x_\text{det}$, $z_\text{det}$, and $t_\text{drift}$, which are then added to the $x$- and $z$-coordinates of the measured cluster positions in order to determine the original undistorted location.



The magnetic field map was simulated with TOSCA code. Hall probe measurements at selected locations verify that the simulation is accurate to within $0.25\%$ when field strength at the center is \SI{0.5}{T}~\cite{Sato2013,Kobayashi2013}. The electric field is calculated by solving Poisson's equation numerically with the Jacobi method~\cite{Jacobi1993}, provided that the electric charge density from space charge and the boundary conditions of the field are known. The field cage configuration described in Section~\ref{exp_set_up} indicates that all six sides can be described by Dirichlet boundary conditions, with the potential on the four vertical sides increasing linearly with $y$-coordinate, and the top and bottom sides having uniform potential.

\begin{figure*}[!h]
\centering
\includegraphics[width=1\linewidth]{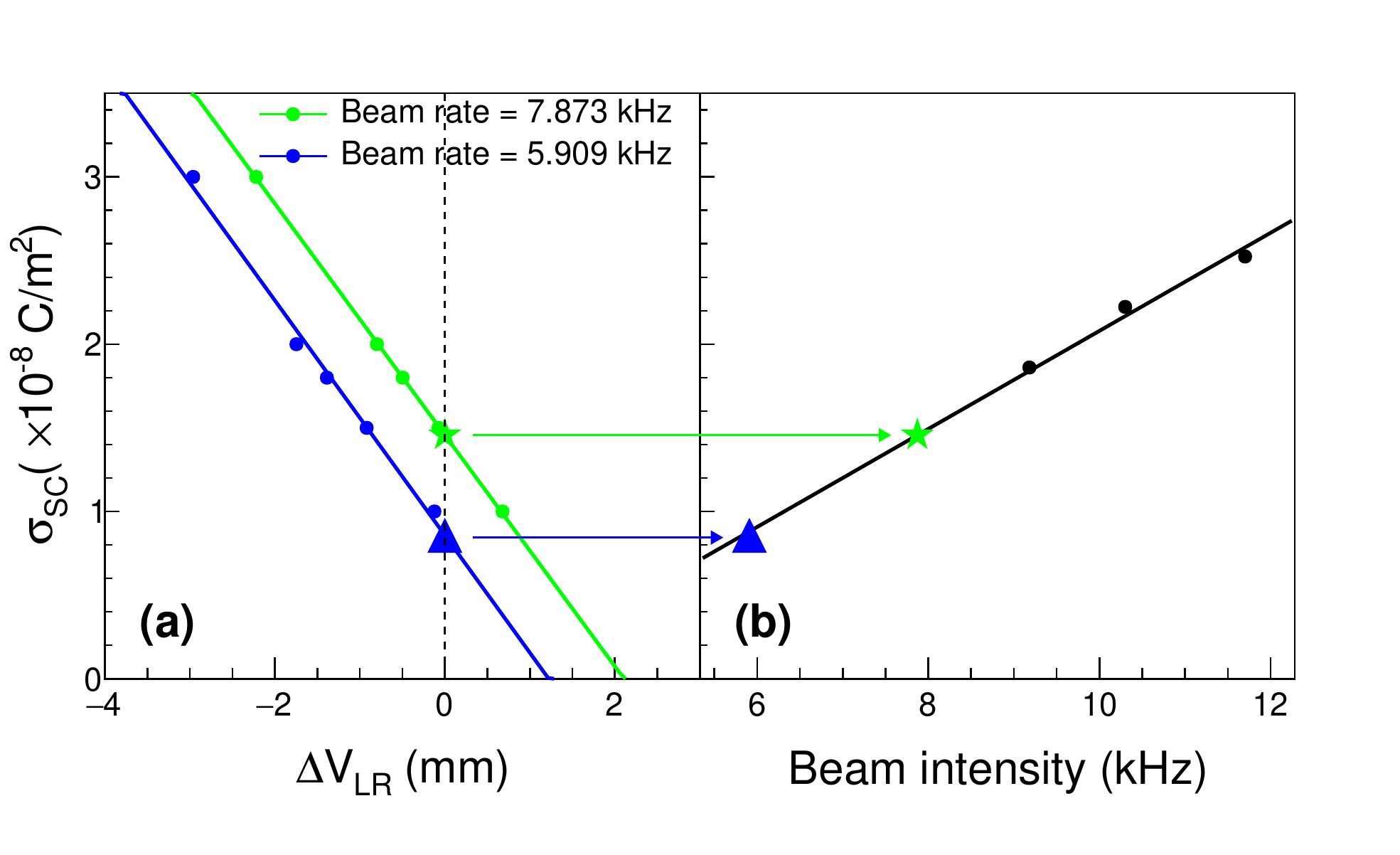}
\caption{(a) $\Delta V_\text{LR}$ after applying the inverse map is plotted against the corresponding $\sigma_\text{SC}$. The two lines corresponds to two runs with different beam intensities. The star and triangle symbols indicate the $\sigma_\text{SC}$ values corresponding to $\Delta V_\text{LR} = 0$. (b) optimal $\sigma_\text{SC}$ for 5 different runs, including the two shown in (a), plotted against their respective run-averaged beam intensity. The star and triangle symbols in (a) and (b) correspond to the same run.}
\label{SC_Correction}
\end{figure*}

The sheet-like space charge distribution, as sketched in Fig.~\ref{SCGraph}, is represented by the charge density $\sigma_\text{SC}$. Without a direct measurement of $\sigma_\text{SC}$, we determine it empirically by scanning a range of values; for each run, five inverse maps are calculated with five trial space charge densities. The value of $\Delta V_\text{LR}$ is calculated and plotted against their corresponding $\sigma_\text{SC}$ in Fig.~\ref{SC_Correction}a. The two example runs shown with different beam intensities exhibit linearity between $\Delta_{\textrm{LR}}$ and $\sigma_{\textrm{SC}}$. Fitting the relationship between $\Delta_{\textrm{LR}}$ and $\sigma_{\textrm{SC}}$ and taking the vertical intercept provides the optimal value of $\sigma_{\textrm{SC}}$.

Ideally, one should optimize $\sigma_\text{SC}$ for each run individually, but it is very computationally expensive. The optimal values of $\sigma_\text{SC}$ with the two runs in Fig.~\ref{SC_Correction}a and three additional runs are plotted against their respective beam intensities in Fig.~\ref{SC_Correction}b, showing a clear linear dependence within the beam variation. The linear approximation of the optimal value of $\sigma_\text{SC}$ as a function of beam intensity reduces the need to optimize $\sigma_\text{SC}$ for each run. 

\begin{figure}[!h]
\includegraphics[width=1\linewidth]{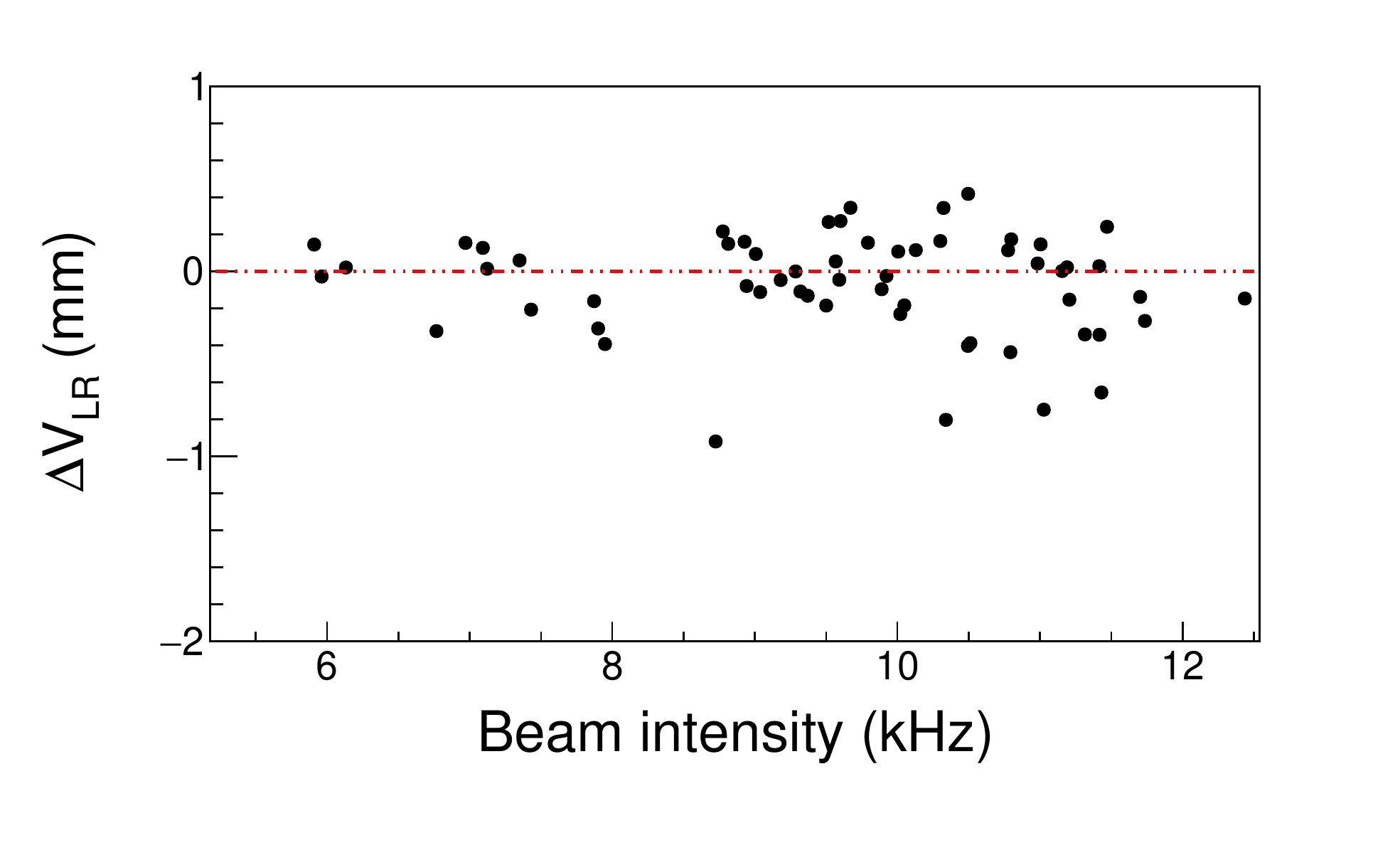}
\caption{$\Delta V_\text{LR}$ vs. beam intensity for all runs after space charge correction.}
\label{SC_BF}
\end{figure}

The dependence of $\Delta V_\text{LR}$ on beam intensity after applying space charge corrections is shown in Fig.~\ref{SC_BF} for all runs.  The reconstructed values of $\Delta V_\text{LR}$ fluctuate around $\Delta V_\text{LR} = 0$, the expected value if the distortion is corrected perfectly.

\begin{figure}[!h]
\begin{subfigure}{1\linewidth}
\vspace*{30pt}
 \includegraphics[width=1\linewidth]{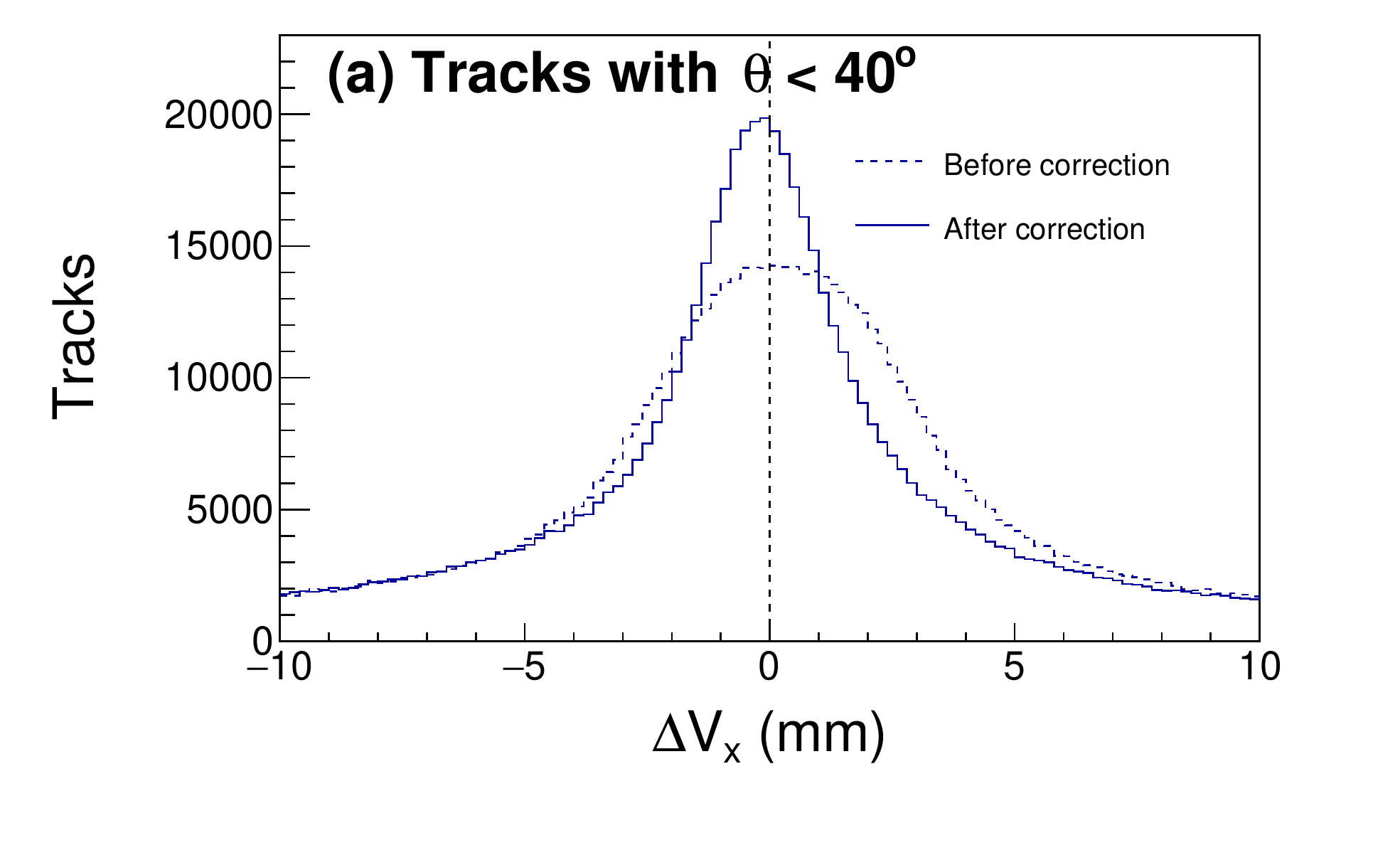}
\end{subfigure}
\begin{subfigure}{1\linewidth}
 \includegraphics[width=1\linewidth]{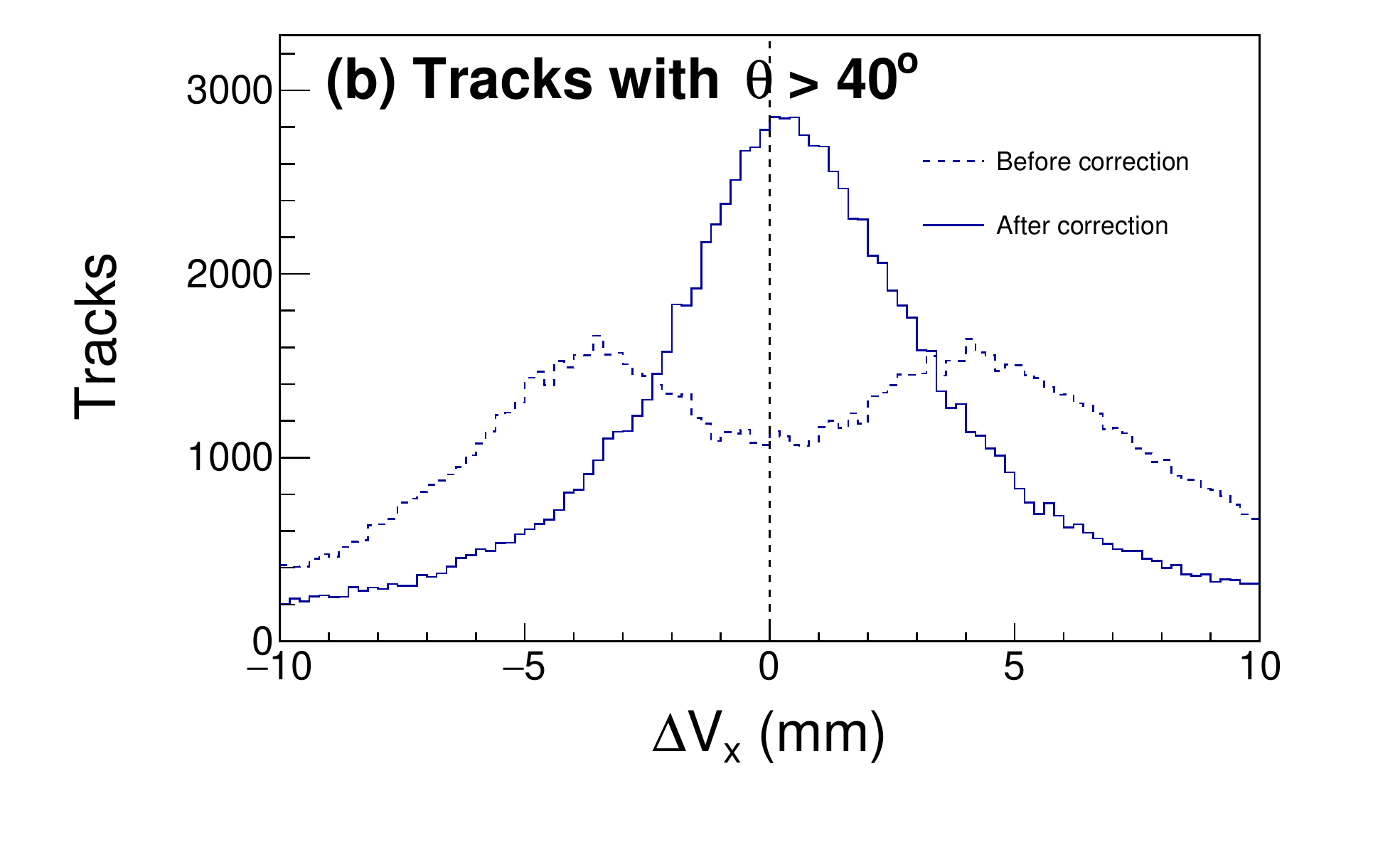}
\end{subfigure}
\caption{$\Delta V_{x}$ distributions before (dashed histogram) and after (solid histogram) applying the calculated inverse map for (a) tracks with $\theta < 40\degree$ and (b) tracks with $\theta > 40\degree$.}
\label{BDCP_aftercor}
\end{figure}

Fig.~\ref{BDCP_aftercor} shows an improvement in the $\Delta V_{x}$ distributions for all tracks with dashed and solid histograms representing data before and after applying the inverse map correction respectively. A significant improvement is also seen in the $\Delta V_{x}$ distributions for tracks with $\theta<40\degree$, seen in Fig.~\ref{BDCP_aftercor}a even though they are minimally affected by the space charge. After the space charge correction, the peak is closer to zero and the full width at half maximum of the distribution improves from \SI{6.6}{mm} to \SI{4.0}{mm}. The improvement is more prominent for tracks with $\theta>40\degree$ as shown in Fig.~\ref{BDCP_aftercor}b, where the bimodal distribution is reduced to a unimodal distribution. The peaks of both distributions in Figs.~\ref{BDCP_aftercor}a and ~\ref{BDCP_aftercor}b deviate from zero by less than \SI{1}{mm}, which is consistent with the uncertainties in the BDC vertex determination.

\begin{figure*}[!h]
\begin{subfigure}{0.5\linewidth}
 \includegraphics[width=1\linewidth, trim=1.5cm 1cm 0.9cm 1cm, clip=true]{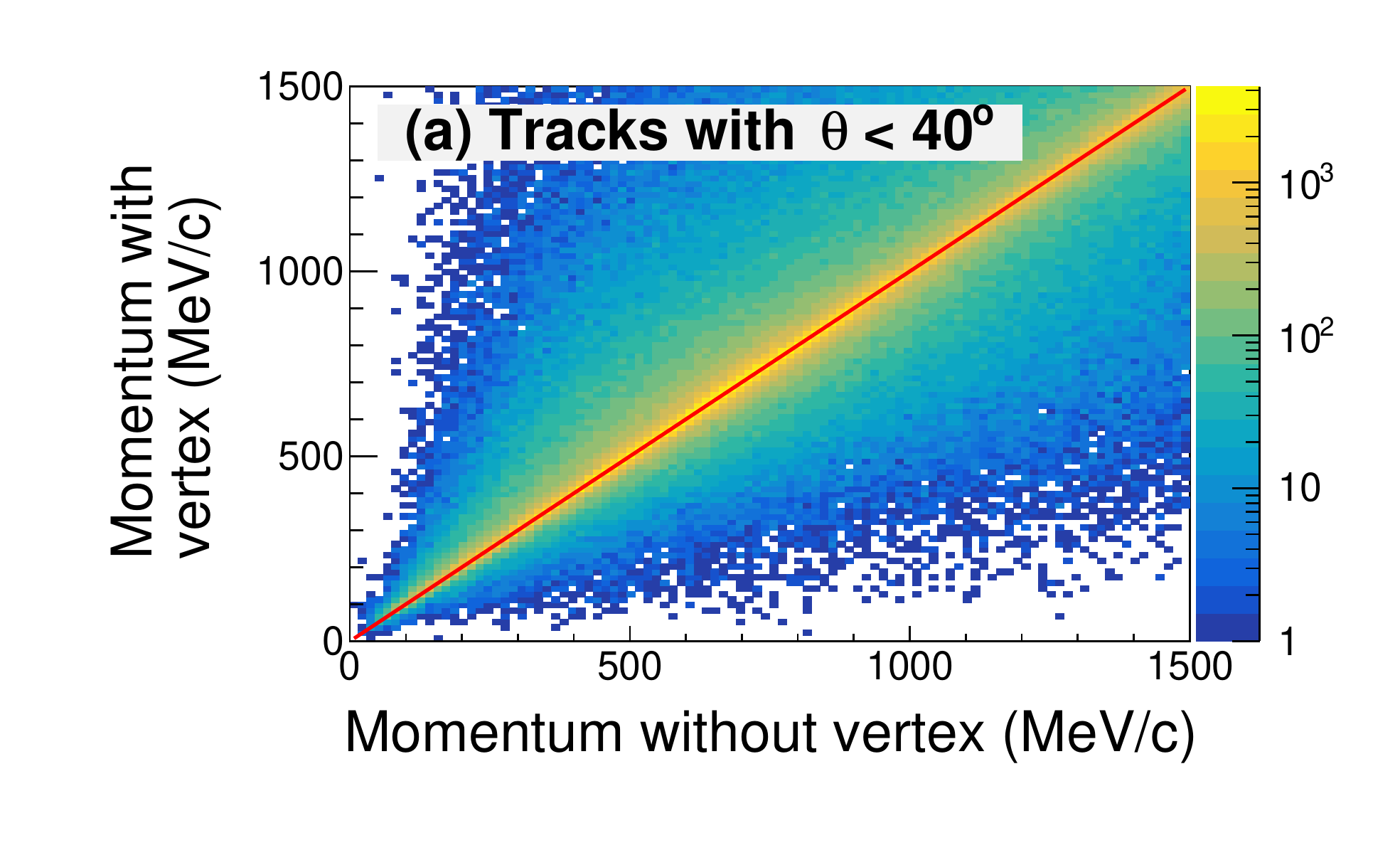}
\end{subfigure}
\begin{subfigure}{0.5\linewidth}
 \includegraphics[width=1\linewidth, trim=1.5cm 1cm 0.9cm 1cm, clip=true]{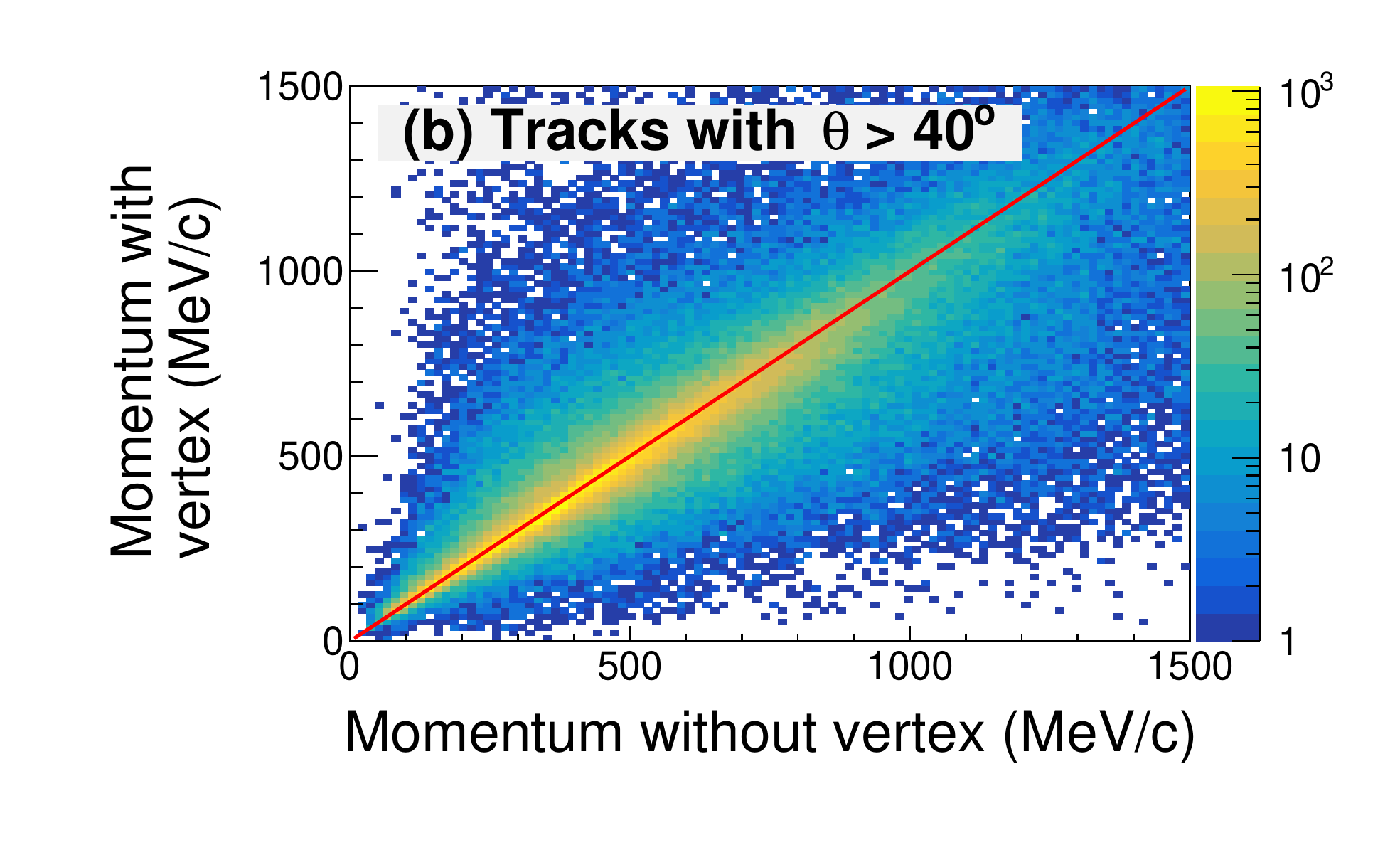}
\end{subfigure}
\begin{subfigure}{0.5\linewidth}
 \includegraphics[width=1\linewidth, trim=1.5cm 1cm 0.9cm 0.5cm, clip=true]{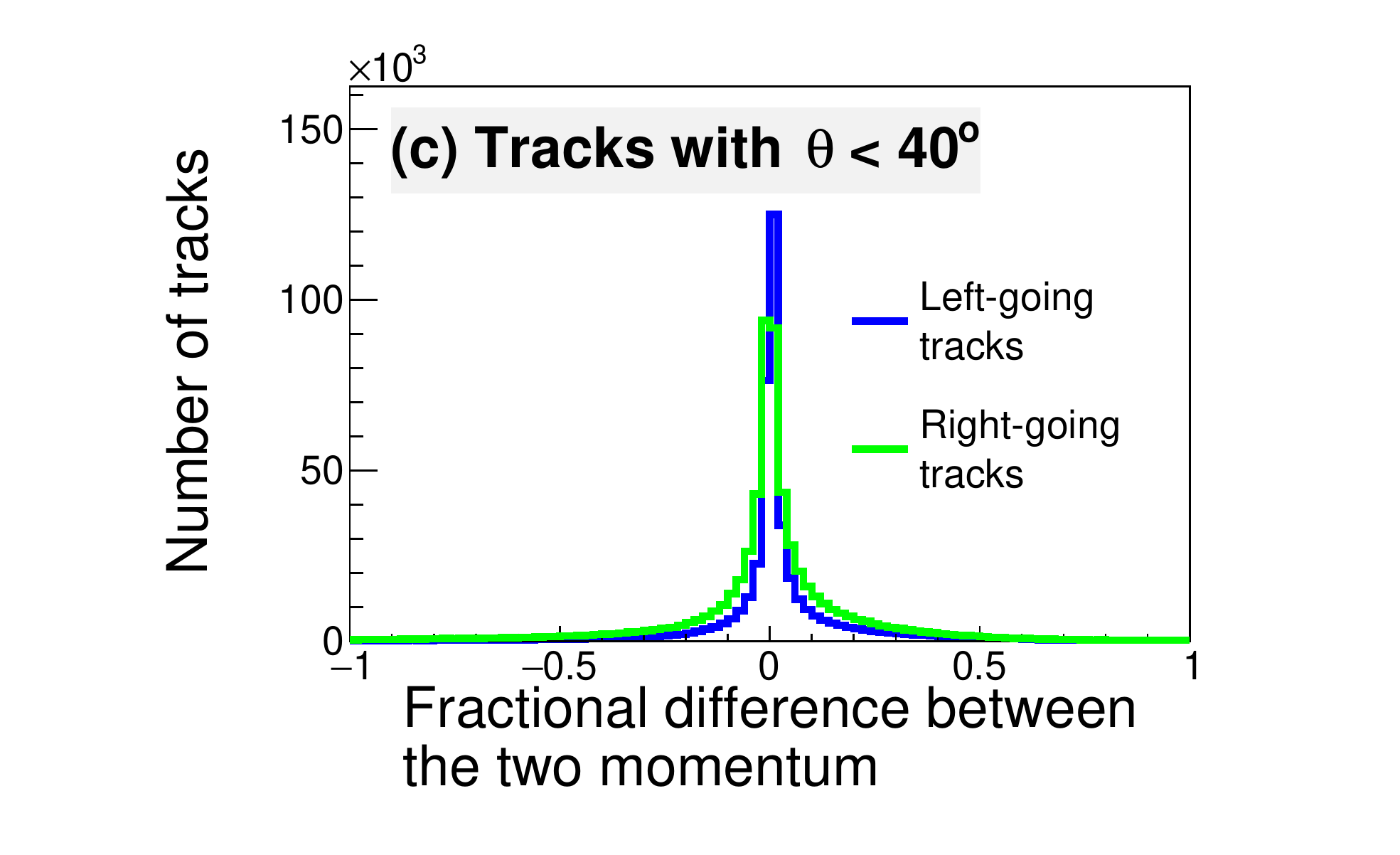}
\end{subfigure}
\begin{subfigure}{0.5\linewidth}
 \includegraphics[width=1\linewidth, trim=1.5cm 1cm 0.9cm 0.5cm, clip=true]{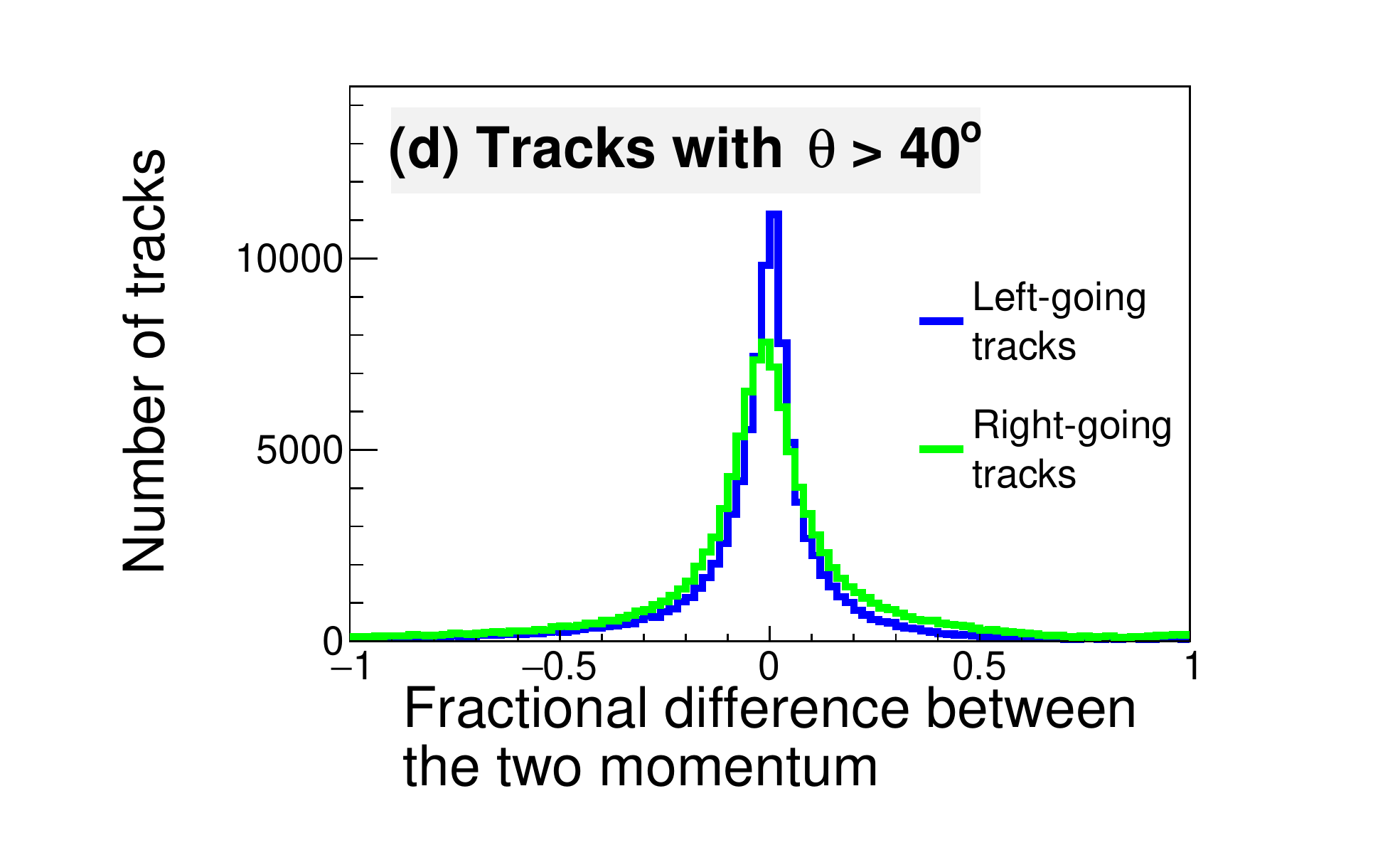}
\end{subfigure}
\caption{The reconstructed momentum using the vertex point plotted against the reconstructed momentum without including the vertex point, for (a) tracks with $\theta$ $<$ 40\degree and (b) tracks with $\theta$ $>$ 40\degree. The color scale corresponds to the number of tracks in each bin. The fractional difference between the two reconstructed momentum is plotted as one-dimensional histogram, for (c) tracks with $\theta$ $<$ 40\degree and (b) tracks with $\theta$ $>$ 40\degree. Distribution from left-going tracks and right-going tracks are overlaid in (c) and (d).}
\label{BDC_P_aftercor}
\end{figure*}

The correction also improves the precision of the reconstructed momentum. This is demonstrated by comparing Fig.~\ref{BDC_P_aftercor}c and~\ref{BDC_P_aftercor}d to Fig.~\ref{BDCP}c and~\ref{BDCP}d. The results are summarized in Table~\ref{BDC_P_summary}. One obvious improvement is the most probable values of the momentum differences are very close to zero after correcting for space charge effects. The full width at half maximum of the distributions of fractional differences between momentum reconstructed with and without vertex are significantly reduced after applying the correction independent of left- or right-going tracks.

\begin{table*}[ht]
\centering
\caption{Summary information for distributions of fractional differences between momentum reconstructed with and without vertex information in Figs.~\ref{BDCP}c, ~\ref{BDCP}d, ~\ref{BDCP_aftercor}c and ~\ref{BDCP_aftercor}d.  }
\begin{tabular}{c c | c c | c c}
\hline\hline
Track direction & Angular cut & \multicolumn{2}{c|}{Most probable value} & \multicolumn{2}{c}{Full width at half Max.} \\ [0.5ex]
 & & Before Cor. & After Cor. & Before Cor. & After Cor. \\
\hline
Left & $\theta < 40\degree$ & $0.006(2)$ & $0.002(2)$ & 0.016(3) & 0.012(3) \\
Right & $\theta < 40\degree$ & $-0.006(2)$ & $-0.002(2)$ & 0.048(3) & 0.028(3) \\
Left & $\theta > 40\degree$ & $0.046(2)$ & $0.002(2)$ & 0.160(3) & 0.076(3) \\
Right & $\theta > 40\degree$ & $-0.142(2)$ & $-0.006(2)$ & 0.260(3) & 0.164(3) \\
\hline
\end{tabular}
\label{BDC_P_summary}
\end{table*}

\section{Discussions and Conclusions}

We have studied the effect of field distortions due to space charge effects in the S$\pi$RIT TPC. The field distortions cause a lateral displacement of drift electrons, which causes the reconstructed tracks to be misaligned from the vertex position and impacts the accuracy of particle momentum determination, especially for particles emitted at large polar angles. 
By solving the Langevin equation (Eq.~\eqref{drift}), we restore the cluster positions back to their original location. In this procedure, the space charge density $\sigma_\text{SC}$ is treated as a free parameter, determined by minimizing the separation between the peaks in the $\Delta V_{x}$ distribution from left-going and right-going particles. This correction should significantly improve the accuracy of the reconstructed momentum, especially for particles emitted at large polar angles. 

This method has a few limitations. Firstly, the space charge sheet density is assumed to be the average value over a run and beam intensity fluctuations that occur within a run ($\sim$30 minutes) are not considered. Another limitation of the method comes from the dependence of charge density on beam type. We found that the relationship between charge density and beam rate was unique to each secondary beam. The charge density depends on the isotopic composition of the secondary beam, which depends on the primary beam settings and beam tuning.
By implementing a laser system in future experiments, this correction could be verified with a direct measurement of space charge distortion. 

Monte Carlo simulation can be performed to further our understanding of the space charge effect beyond our current observables and to quantify the improvement in determining momentum resolution and accuracy. An effort is underway to simulate the detector response of the S$\pi$RIT TPC.

\section{Acknowledgments}

This work was supported by the U.S. Department of Energy under grant Nos. DE-SC0014530, DE-NA0002923, US National Science Foundation under grant No. PHY-1565546, the Japanese MEXT KAKENHI (Grant-in-Aid for Scientific Research on Innovative Areas) under grant No. 24105004 and the National Research Foundation of Korea under grant Nos. 2016K1A3A7A09005578, 2018R1A5A1025563. The computing resources for analyzing the data was supported by the HOKUSAI-GreatWave system at RIKEN, the Institute for Cyber-Enabled Research (ICER) cluster at Michigan State University, and the EMBER cluster at the NSCL.

\bibliographystyle{elsarticle-num}

\end{document}